\documentclass[runningheads]{llncs}
\usepackage{amssymb}
\usepackage[usenames,dvipsnames,svgnames,table]{xcolor}
\usepackage{amsmath}
\usepackage{graphicx}
\usepackage{hyperref}
\usepackage{color}
\usepackage{epstopdf}
\usepackage{cleveref}
\usepackage[svgnames]{xcolor}
\usepackage{braket}
\usepackage{cite}
\usepackage{multirow}
\hypersetup{hidelinks,colorlinks=true,allcolors=DarkBlue}

\usepackage[normalem]{ulem}

\usepackage{enumitem}
\usepackage[ruled, vlined]{algorithm2e}

\usepackage[normalem]{ulem}

\begin{document}
\title{Towards security recommendations for public-key infrastructures for production environments in the post-quantum era}
\author
{Sergey E. Yunakovsky$^{1,2}$\and
Maxim Kot$^{1,2}$\and
Nikolay Pozhar$^{1,2}$\and
Denis Nabokov$^{1,2}$\and \\
Mikhail Kudinov$^{1,2}$\and
Anton Guglya$^{1,2}$\and 
Evgeniy O. Kiktenko$^{1,2}$\and \\
Ekaterina Kolycheva$^{3}$\and
Alexander Borisov$^{3}$\and
Aleksey K. Fedorov$^{1,2}$}

\authorrunning{S.E. Yunakovsky et al.}
\titlerunning{Post-quantum security}
\institute{$^1$QApp, Skolkovo, Moscow 143025, Russia \\ 
$^2$Russian Quantum Center, Skolkovo, Moscow 143025, Russia \\ 
$^3$Bosch Corp. Sector Res. \& Adv. Eng., Saint Petersburg 198095, Russia \\
\email{akf@rqc.ru}}

\maketitle
\begin{abstract}
Quantum computing technologies pose a significant threat to the currently employed public-key cryptography protocols.
In this paper, we discuss the impact of the quantum threat on public key infrastructures (PKIs), which are used as a part of security systems for protecting production environments.
We analyze security issues of existing models with a focus on requirements for a fast transition to post-quantum solutions. 
Although our primary focus is on the attacks with quantum computing, we also discuss some security issues that are not directly related to the used cryptographic algorithms but are essential for the overall security of the PKI.
We attempt to provide a set of security recommendations regarding the PKI from the viewpoints of attacks with quantum computers.  
\end{abstract}

\section{Introduction} \label{intro}

In the digital era, cryptography plays a central role in ensuring the security and privacy of communications, which are crucial for various fields ranging from personal data to critical infrastructure.
Cryptographic techniques are used throughout government and industry to authenticate the source and protect the confidentiality and integrity of information.
Existing cryptographic tools substantially use the concept of public-key cryptography.
It is a technique that enables entities to securely communicate on an insecure public network by solving the key distribution problem, and reliably verify their identities via digital signatures. 
Public-key cryptography is also known as asymmetric cryptography since the parties of communications use two types of keys: 
Public keys, which may be known to others, and private keys, which may never be known by any except its owner.
This is its important difference in compare with symmetric cryptography, which relies on the use of the only one secret key shared between the parties (however, the problem of key distribution for symmetric cryptography is challenging; see below).

In its turn, public-key cryptography forms a basis for a public key infrastructure (PKI), which is a set of roles, policies, hardware, software, 
and procedures needed to establish compliance between real-world parties of communications (like people, manufacturers, or devices) and public keys. 
Certificates are basic digital documents that state the correspondence between an entity and its public key~\cite{AdamsLloyd2002}.
PKI plays a crucial role in protecting many processes and, in particular, all phases of product development and distribution in production environments. 

Security of public-key cryptography, which defines the security of PKIs, relies on the concept of NP problems, which have proof verifiable in polynomial time.  
For example, multiplying two large prime numbers is computationally easy (at it is then easy to correct that multiplication of two prime numbers gives the correct integer number), 
but finding the prime factors of a given product is hard --- it can take a conventional computer thousands years to solve for large numbers. 
In terms of public-key cryptography, this means that the key distribution problem (signing documents and checking the signature using the public key) is computationally easy, whereas obtaining a private key with the known public key is computationally hard. 
NP problems, such as integer factorization and discrete logarithm problems, are used in modern cryptosystems Rivest-Shamir-Adleman (RSA) cryptosystem~\cite{RSA1978} and Diffie-Hellman scheme~\cite{DH1976}, correspondingly. 
Under the assumption that existing computers could not solve these mathematical tasks in a reasonable time, modern public-key cryptography techniques, such as RSA and Diffie-Hellman schemes, seem to be secure.

A new generation of computing devices, which use operate on the principles of quantum physics, so-called quantum computers, would allow solving various mathematical tasks much faster than their classical counterparts.
Examples of such tasks include NP problems, which are behind the security of mentioned above RSA and Diffie-Hellman schemes, with the use of quantum Shor's algorithm~\cite{Shor1997}.
In practice, this means that an adversary with a quantum computer will be able to obtain a private key from a corresponding public key.  
Consequently, quantum computers with enough computing power (so-called quantum volume) would allow breaking popular and widely deployed tools for cryptographic protection.
Quantum computing also has an impact on symmetric cryptography since quantum Grover's algorithm~\cite{Grover1996} provides a quadratic speed-up in the brute force search, but this is not dramatic.  
Thus, quantum computing poses a threat to currently used information security protocols based on PKI, in particular those used in the Transport Layer Security (TLS),
which is the security protocol behind the Hypertext Transfer Protocol Secure (HTTPS)~\cite{Digicert}. 

However, not all existing security tools are vulnerable to attacks with quantum computers~\cite{Wallden2019,Bernstein2017}.
Currently, serious efforts are concentrated on developing quantum-resistant cryptographic tools and the strategy of their deployment to the currently existing infrastructure. 
A number of cryptographic systems, which use these methods, 
are considered as candidates in the National Institute of Standards and Technology (NIST) Post-Quantum Cryptography Standardization and by European Telecommunications Standards Institute (ETSI). 

The deployment of quantum-resistant solutions are of significant importance for many information systems. 
Here we focus on large-scale production environments, where most of the security tools for protecting supply chains, distribution networks, financial management systems and communications, 
and control systems are based on public key infrastructure (PKI)~\cite{AdamsLloyd2002,Landrock2006,Hoglund2020,Yong2006,Hanke2007}.
We use the typical structure of PKI of a production environment, which is provided by Bosch and presented below, as an example for analysis from the viewpoint of potential attacks with quantum computing. 
To avoid major losses~\cite{Mosca2017}, companies and firms that substantially use PKI should pay attention to the quantum threat and create a post-quantum security strategy. 

In this work, we consider the impact of the quantum threat on PKI, which is used for protecting production environments.
We analyze the security issues of the model of injecting the trusted certificate and provide security recommendations regarding the PKI from attacks with quantum computers.
Although real-world production environments are frequently considered as a subject of the analysis from the viewpoint of upcoming threats from quantum computing technologies,
our work (to the best of our knowledge) demonstrates the first detailed holistic consideration of strategic changes in PKI for providing post-quantum security.
We also discuss the applicability of post-quantum algorithms in the security systems for production environments.

Our paper is organized as follows.
In Sec.~\ref{sec:impact} we analyze an impact of quantum computers on modern cryptographic tools.
In Sec.~\ref{sec:analysis} we consider quantum security of state-of-the-art PKI model for a production environment.
In Sec.~\ref{sec:schemes} we discuss the applicability of post-quatnum algorithms.
In Sec.~\ref{sec:recommendations} we form security recommendations that are based on our analysis. 
We conclude in Sec.~\ref{sec:conc}.

\section{Impact of quantum computers on cryptography}\label{sec:impact}

Here we briefly review the state-of-the-art in cryptoanalysis with the use of quantum computers (for a review, see Ref.~\cite{Mavroeidis2018}).
Then we consider existing options for protecting PKI in the post-quantum era.

\subsection{Quantum threat for cryptography}

\subsubsection{Symmetric cryptography.}
Cryptography implies various techniques, which can be divided into two large categories: symmetric (private-key cryptography) and asymmetric (public-key cryptography). 
Symmetric cryptographic techniques use the same key for encryption and decryption processes. 
Symmetric cryptography is fast, relatively easy to implement and operate, but it suffers from two main difficulties.
The first is the issue of the confidential key distribution between distinct parties.
Symmetric cryptography is still widespread among some organizations that use, for example, trusted couriers for the key distribution that is indeed complicated in the era of digital communications.
The second problem is the need to change keys quite frequently to reduce the probability of discovering keys by an attacker. 
Therefore, symmetric cryptographic techniques are useful only under the condition of having an efficient method for distribution and changing keys.

Quantum computers have an impact on symmetric cryptographic primitives, but exponential speedups in their cryptanalysis are not expected. 
Grover's algorithm would allow quantum computers a quadratic speedup in brute force search~\cite{Grover1996}.
Then the key management in terms of the key size and the key refresh time for such primitives needs to be reconsidered. 
For example, AES-256 is considered quantum-secured with 128 bits of quantum security (in the view of quadratic speedup in brute force search).

\subsubsection{Public-key (asymmetric) cryptography.}
The situation differs for the currently deployed public-key (asymmetric) cryptography tools, which use a pair of public/private keys. 
Public-key cryptographic primitives are mainly mathematical problems that are believed to be computationally hard.
They are used as the basis in popular cryptographic schemes such as RSA, Diffie-Hellman, ECDSA (Elliptic Curve Digital Signature Algorithm), etc~\cite{Wallden2019}.

However, quantum computers can solve the problems, which are behind the security of these primitives in polynomial time using Shor's algorithm~\cite{Shor1997}. 
The question of the required resources from the side of quantum computers for factoring integers and computing discrete logarithms in finite fields with the use of Shor's algorithm~\cite{Shor1997} 
is a subject of extended research activities~\cite{Shor1997,Griffiths1996,Zalka2006,Fowler2012,Ekera2017,GidneyFowler2019,Gidney2019}.
The one of latest result~\cite{Gidney2019} is the scheme that uses $3n+0.002n\lg{n}$ logical qubits (i.e. qubits wokring without errors), $0.3n^2+0.0005n^3\lg{n}$ Toffoli gates, and $500n^2+n^2\lg{n}$ measurement depth to factor $n$-bit RSA integers.
This means that 2048 bit RSA integers can be factorized in 8 hours using 20 million noisy qubits~\cite{Gidney2019}, whereas one of the largest existing gate-based quantum computers has about 53 noisy qubits~\cite{Martinis2019}. 
Alternative proposal is to use a computing protocol with a multimode memory, which allows factoring 2048 RSA integers in 177 days with 13436 qubits~\cite{Gouzien2021}.
Thus, current quantum computers are far from being capable of executing Shor's algorithms for cryptographically relevant problem sizes~\cite{Gidney2019}.
There is an increasing interest in alternative schemes for quantum factoring, such as variational quantum factoring~\cite{Aspuru-Guzik2019}.
Variational quantum factoring is an alternative to Shor's algorithm, which employs established techniques to map the factoring problem to the ground state of an Ising Hamiltonian. 
It starts by simplifying equations over Boolean variables in a preprocessing step to reduce the number of qubits needed for the Hamiltonian. 
The examination of a more detailed analysis of the potential scalability of such an approach using realistic noisy intermediate-scale quantum devices is under investigation~\cite{Aspuru-Guzik2019}.

Thus, the existence of Shor's algorithm makes the corresponding public-key cryptography methods vulnerable. 
Therefore, most of the existing and currently used primitives used in PKI should be replaced to guarantee security against quantum attacks.  
In this case, it is not enough to reconsider the key size --- these algorithms should be replaced as soon as they are no longer secure. 

\subsubsection{Store now -- decrypt later.}
One of the most important existing problems is related to the so-called "store now -- decrypt later" attack.
The idea is that the adversary is harvesting data in encrypted form, in the hope that quantum computing will help them to uncover valuable information from it in the future.
That is why for some particular applications dealing with long-term sensitive information, one should think about the priority replacement of cryptographic primitives on quantum-secured ones. 
This fact is expressed in Mosca's theorem says, which states the following: 
We need to start worrying about the impact of quantum computers when the amount of time that we wish our data to be secure for ($X$) is added to the time it will take for our computer systems to transition from classical to post-quantum 
($Y$) is greater than the time it will take for quantum computers to start breaking existing quantum-susceptible encryption protocols ($Z$).

Importantly, this paradigm can be extended to the idea of cryptographic agility (crypto-agility),
which is the capacity for information security systems to switch on alternatives to the original encryption method or cryptographic primitive without significant change to system infrastructure.  
In the terms of Mosca's theorem this requires to the minimization of the transition time to quantum resistant solutions. 

\subsection{Quantum-resistant cryptography}

There are several ways to protect information infrastructure in the era of quantum computers, the so-called post-quantum era~\cite{Wallden2019}.
The crucial problems, which are typically solved using public-key cryptography primitives, are related to the key distribution problem and digital signatures. 
There exist several practical ways of solving these problems in the post-quantum era. 

\subsubsection{Quantum key distribution.}
The first is to replace public-key cryptography with quantum key distribution (QKD, also known as quantum cryptography), which is a hardware solution based on transmitting information using individual quantum objects~\cite{Gisin2002}. 
The main advantage of this approach is that the security relies not on any computational assumptions but the laws of quantum physics. 
The idea of QKD is that two legitimate users (Alice and Bob) have the pre-shared authentication key and the communication channel. 
Then they establish a QKD protocol that allows them to obtain a raw quantum key, which contains some errors and some information about the key that is potentially known to the adversary. 
In the QKD security proofs, it is assumed that all errors in raw quantum keys are due to eavesdropping~\cite{Gisin2002}. 
Alice and Bob initiate the post-processing procedure using the authenticated public channel. 
As a result, Alice and Bob have a key for applications, and it is proven to be information-theoretically secure against arbitrary attacks, including the quantum ones~\cite{Scarani2009}. 
QKD-generated keys can be used for conventional symmetric encryption, such as AES, and used to frequently refresh keys.

Remarkable progress in the deployment of several quantum key distribution networks around the globe has been performed. 
Various industry cases of QKD use, such as those in finance, telecommunications, and data center infrastructure, have been demonstrated~\cite{IDQ,QRate}. 
The largest QKD network is by now deployed in China, which spans 4600 km and includes the link between the cities of Shanghai, Hefei, Jinan, and Beijing and a satellite link spanning 2600 km between two observatories~\cite{Pan2021}. 
The operation of such QKD networks requires the use of trusted relay nodes because of the presence of optical losses in communication channels, limiting the distance for the realization of the QKD protocol. 

At the current stage, QKD technology faces several challenges~\cite{Gisin2002},
which makes it best suitable for some domain-specific applications, such as the protection of highly-loaded communications links at a distance, which does not require the use of intermediate nodes~\cite{Lo2014,Lo2016}. 
We note that the practical implementation of digital signatures based on quantum key distribution in the industrial environments seems to be quite complicated from the practical point of view. 

\subsubsection{Post-quantum cryptography.}
An alternative way to guarantee the security of communications is to switch to a new type of public-key cryptosystems. 
Fortunately, not all public-key cryptosystems are vulnerable to attacks with quantum computers~\cite{Bernstein2017}. 
Several cryptosystems for key distribution and digital signature, which strive to remain secure under the assumption that the attacker has a large-scale quantum computer, have been suggested. 
These schemes are in the scope of so-called post-quantum cryptography. 
Post-quantum protocols are based on different mathematical approaches, such as
the shortest vector problem in a lattice~\cite{Regev2009, Hanrot2007, Micciancio2002}, 
learning with errors~\cite{Regev2010, Lyubashevsky2010, FrodoKEMSubmission, CRYSTALS-KYBERSubmission, Albrecht2015, Kirchner2015, Arora2011, Schnorr1994, Chen2011, NewHopeSubmission,DilithiumSubmission}, 
solving systems of multivariate quadratic equations over finite fields~\cite{Patarin1996,Faugere2003,Beullens2017,GeMSSSubmission,LUOVSubmission,RainbowSubmission}, 
finding isogenies between elliptic curves~\cite{Jao2011, Costello2017, Costello2016, Koziel2017, Steven1999, Delfs2016, Zhang2005, Tani2007}, 
decoding problems in an error-correcting code~\cite{Berlekamp1978, Alekhnovich2003, May2015, Becker2012, Bernstein2010, Drucker2020,RFC_XMSS,RFC_LMS},
security properties of cryptographic hash-functions~\cite{Buchmann2011, SPHINCS+Submission, Hulsing2016, Bernstein2019, Rogaway2004,Sphincs+Comment}, and other primitives~\cite{PICNICSubmission}.

\subsubsection{Hybrid quantum-secured cryptography.}

A useful strategy is the combination of different cryptographic techniques~\cite{Fedorov2017}. 
For example, one can combine QKD with symmetric encryption or with post-quantum cryptography, where the latter can be used for various purposes (e.g. for authentication purposes in QKD protocol~\cite{Pan2020}). 
In addition, a hybrid quantum-secured infrastructure may use QKD for protecting highly-loaded communications link at the distance, 
which do not require the use of intermediate nodes, whereas end-users without direction connections can be protected by means of post-quantum cryptography. 

\subsubsection{Standardization processes.}
Both quantum and post-quantum cryptography undergo active standardization processes.
In particular, standardization of the QKD technology is considered by several agencies, such as ETSI and ITU.

The standardization of the post-quantum cryptography currently is centered around the NIST initiative~\cite{NISTcomp}, 
which are intended to choose and standardize post-quantum algorithms for stateless digital signatures and key encapsulation mechanisms/public key encryption.
The process is similar to the previous hash function and AES NIST competitions.
Up to date, two rounds have already finished~\cite{StatusReport} and the third round is in progress.
The final third round should result in a choice of algorithms for standardization.

\section{Analysis of quantum security of state-of-the-art PKI model for a production environment} \label{sec:analysis}

PKI is a set of measures that are needed to use digital certificates and manage public-key encryption~\cite{AdamsLloyd2002,Landrock2006,Hoglund2020,Yong2006,Hanke2007}. 
The main goal of PKI is to bind entities with public keys of asymmetric cryptosystems. 
The binding is established with the use of certificates. 
A certificate is a dataset that gives information about the entity and its public key. The certificate is signed by a trusted third party, whose public key is known. 

The core idea of the PKI is to achieve the root of trust during all phases of the product development and distribution.
That is why it is important to implement the key hierarchy and protection of the data in rest to guarantee the PKI resistance against various possible threats.
Additionally, an efficient PKI model should contain mechanisms for the control of already enrolled certificates and keys in a way that allows revocating keys and detecting the compromise of the particular parts of the system.
On the basis of widely used assumptions we can separate the PKI tasks in the following way:
\begin{enumerate}
	\item {\bf enrollment and provision} of new certificates;
	\item {\bf authentication and verification} of involved parties and certificates;
	\item {\bf revocation and detection} of compromised or expired certificates.
\end{enumerate}

Currently used PKI schemes are mostly based on non-quantum-resistant cryptographic mechanisms.
This section aims to analyze the state-of-the-art PKI model for formulating security recommendations. 
In the underlying sections, we describe security aspects for each of the listed functional parts.

\subsection{PKI model for production environments}

\begin{figure}[t]
\centering
\includegraphics[width=1\textwidth]{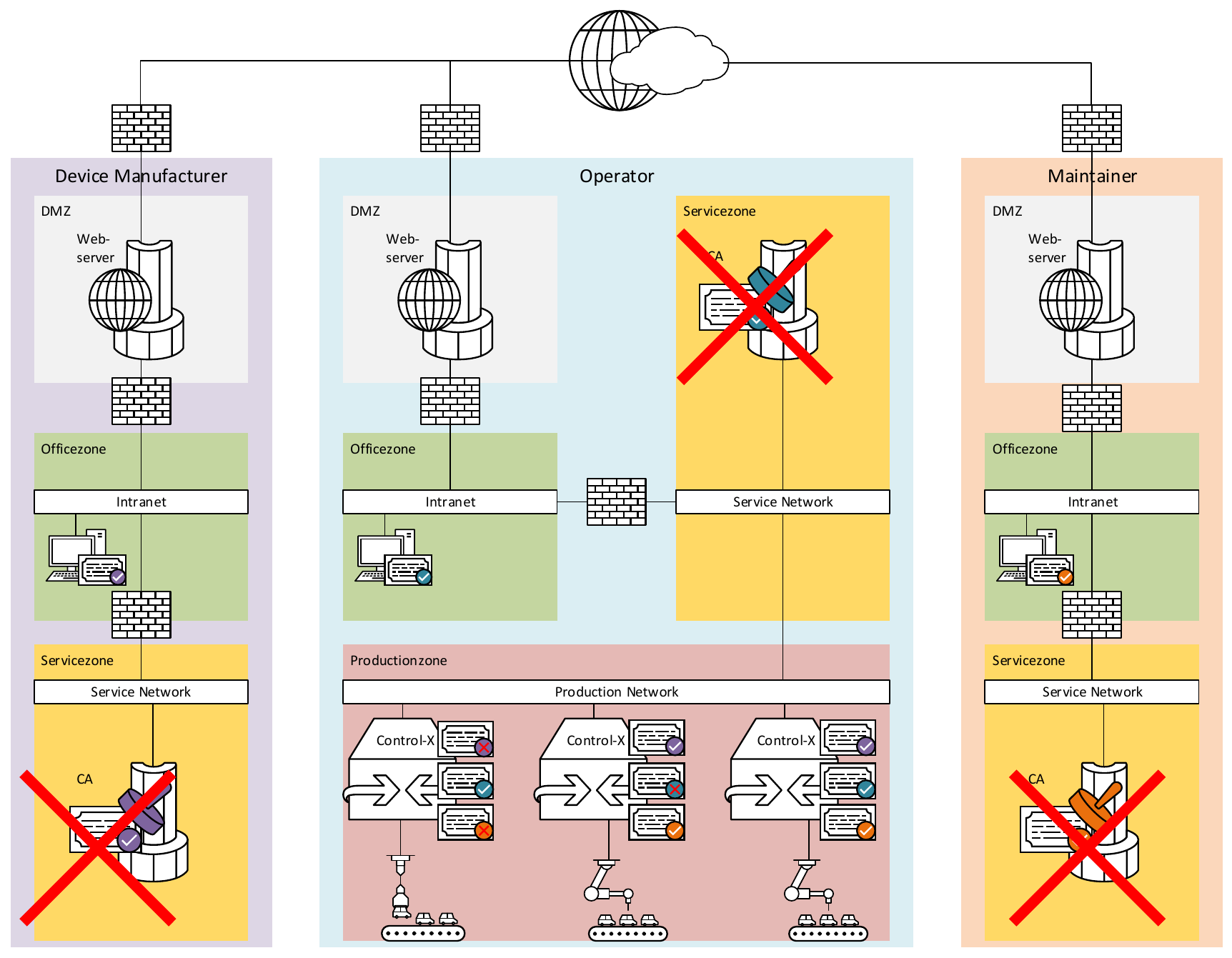}
\vskip -4mm
\caption{PKI structure of a production environment.}
\label{fig:scheme}
\end{figure}

Our further analysis is presented for a specific PKI model, which is used in production environments (we use the concrete scheme, which is provided by Bosch). 
The diagram that described the existing scheme of the certificates enrollment is shown in Fig.~\ref{fig:scheme}.
The main functional goal of this scheme is to inject the trusted certificate into the final product.
In this particular case, the final product is a produced device. 

We use the following assumptions regarding the provided scheme.
\begin{enumerate}
	\item The main certificate authority (CA, PKI Frontend) is considered to be trusted. 
	The compromise of the core CA may lead to the security breach in the PKI regardless of applied security countermeasures. 
	An alternative solution is to develop the PKI model that is based on the decentralized root of trust. 
	However, this topic is outside the context of the present paper as the decentralized PKI requires technologies similar to blockchains (whose cryptographic security is also a subject of research~\cite{Kiktenko2019,Fedorov2018}).
	\item The perimeter of the production zone and service zones are trusted or at least contain mechanisms to notify other parties about derivations of planned activities (e.g. certificate enrollment) from expected behaviour. 
	Such behaviour may be caused by various reasons that include: 
	\begin{itemize}
		\item unauthorized access to the system by the malicious or unauthorized actor;
		\item malfunction of the system caused by environmental conditions, power supply, hardware or software issues;
		\item infection of the system with malware.
	\end{itemize}

	The information regarding the current state of the production and service zones must be handled by the monitoring system, which may efficiently notify authorized parties. 
	Communication channels and threshold values used to detect the compromise must be aligned between parties during the development of the monitoring system. 
	As an example, it is not possible to share the information about the current state of the service zone using the same communication channel, 
	which is used for the communication with the production line as both of them (including the communication channel itself) may be compromised.
	
	\item The used algorithms at the current stage are compliant against publicly available standards (e.g. NIST FIPS 140-2~\cite{NIST-FIPS-140-2}). 
	The misuse of cryptography modes and parameters may compromise the data in rest regardless of applied countermeasures.
	\item The malicious actor may be one of the following:
	\begin{itemize}
		\item an external party aiming to compromise the confidentiality of data in transit to access the content of the firmware update and device configuration;
		\item device manufacturers, which are not authorized to access proprietary information regarding the internal structure of the device and software; 	
		for example, such a manufacturer may have physical access to one of the devices distributed in the market aiming to perform the reverse-engineering of the device to clone it and create a similar product;
		\item a group of highly experienced specialists in the field of informational security aiming to compromise proprietary information about the production line, company, and products.
	\end{itemize}
	
\end{enumerate}

As soon as we consider a specific example of the currently used scheme of relations between involved parties and the set of business requirements for this scheme, 
we adjust our assumptions based on the provided scheme as follows.
\begin{enumerate}
	\item All parties (manufacturer, maintainer, operator) may want to inject their own certificates, which are not related to a specific PKI model or associated with each other.
	\item The device should be able to generate a certificate by itself.
	\item All parties may use one of the following mechanisms to inject certificate:
	\begin{enumerate}
		\item the company frontend;
		\item special application programming interface on the device itself;
		\item directly uploads the certificate on the device using the device's API.
	\end{enumerate}
\end{enumerate}
Some additional technical details are placed in Appendix A.

We note the following potential weaknesses in the provided scheme.
\begin{enumerate}
	\item The public network is compromised and anyone can get access to transmitting data. 
	This situation includes eavesdropping and modification of data in transit. Moreover, in some cases, it may be possible to save communication data and decrypt it lately with access to the operable quantum computer.
	\item The provided scheme does not cover the aspect of communications between parties.
	\item The injection process takes place (in the scheme as is) without verification of the device/backend integrity. 
	The device integrity must be achieved through the hardware level isolation (virtualization) technologies and embedded in the protected memory shared with the backend private key and information regarding the device itself 
	(hardware identifies, device specifications). 
	The attestation process may be performed inside the isolated environment of the device to verify its integrity against embedded information. 
	Additionally, the device may integrate various tamper detection techniques, both software and hardware to verify its integrity. 
	The private key stored inside protected memory grants the trustworthiness of the attestation data. 
	The verification of the backend authenticity may be achieved through the verification of shared by the backend certificate within an isolated environment against embedded in the protected memory information.
\end{enumerate}

The manual injection of the certificate is considered to be a 'work around' and is not related to a unified structure provided by the PKI.
Then we build a proposed scheme based on assumption that injection of certificates took place using a company's frontend or the device API. 
Moreover, the usage of the unified method of certificate injection allows describing each of the participants involved in the injection equally. 
In other words, the relations between the manufacturer and the integrator are not taken in place as both of them are seen by the PKI as regular nodes.


\section{Security recommendations}\label{sec:recommendations}

Here we would like to summarize recommendations regarding the overall structure of the PKI with the focus on threats coming from quantum computing. 
We recommend improving the scheme in a number of aspects. 
First, one needs to take into account existing (non-quantum) attacks on PKI schemes. 
Second, it is important to take into account possible risks, which are related to quantum threats. 
These recommendations are a basis for the improvement of security aspects of the final holistic solution for PKI.
Our list of recommendations is as follows:
\begin{itemize}
	
	\item Cryptography in place.
	
		\begin{itemize}
		
		\item CAs certificates and cryptography considered to be unified.
		We assume that all parties sharing the same set of software development kits (SDKs) and software/hardware to perform required cryptography operations. 
		To achieve this, the first step is to enforce universal security requirements for the software. 
		
		\item Software should pass security evaluation and should be developed according to the Security Code Practice.
		
		\item SDKs in this model are assumed to be unified. 
		Then it is possible to improve the security of cryptography operations. 
		For example, it is possible to embed the information regarding the current state of the service zone and used software in the certificate itself to ensure that the state of the CA is trusted. 
		Moreover, the time required for the migration of the architecture to the post-quantum era, in this case, is significantly reduced since one can use the unified mechanism of the software update and deployment.

		\item We recommend using the X.509 format for the certificate. 
		This due to the fact that it supports an extensible scheme of embedded data. 
		It is possible to store multiple public keys from different algorithms in the same certificate. 
		For example, it is possible to embed in the signed certificate both keys RSA key and post-quantum Falcon key. 
		Such an approach allows both supporting existing standards in cryptography and ensuring post-quantum security. 
		However, the rollback protection mechanism must be implemented and enforced to mitigate downgrade attacks against the hybrid scheme.
		
		\end{itemize}
		
	\item Communications.
	
		\begin{itemize}
	
		\item Parties during the communication may operate in different time zones and conditions. 
		Then it is possible for one of the parties to be unavailable during the required time period. 
		A presumable solution for such a challenge is to use limited use certificates with a very short lifetime, which are signed with the private key of the CA. 
		
		\item Attacks with quantum computers are able to completely compromise the PKI model that is based on the usage of a set of algorithms, which are not resistant to quantum attacks. 
		The extensible scheme, which allows one to replaces signing algorithms on-a-fly requires significant changes in the manufacturing cycle (e.g. firmware verification, secure boot, certificates enrollment).

		\item As an additional improvement, it is recommended to develop the PKI model with the possibility to extend a set of used algorithms with the support of post-quantum algorithms and to perform a regular evaluation of the implemented scheme.
		 It should be ensured that the scheme works in a crypto-agile manner. 

		\end{itemize}
		
	\item Enrollment and provision of certificates.
	
		\begin{itemize}
		
		\item The enrollment process is the initial point of the PKI model, so it deserves additional attention before the process of certificate generation can be started. 
		As a consequence, the PKI model should include the trusted channel between parties, which allows parties to ensure their states and initializing the enrollment process. 

		\item We do not recommend using the same channel both for the exchange of certificates (cryptographic materials) and control signals. 
		
		\item We recommend using hardware-backed authentication methods for the critical parts of the enrollment process (e.g. confirmation of the signing of the second level certificate). 
		This can be done with the help of USB tokens or similar solutions.

		\item It is possible also to improve the trustworthiness of CAs. 
		This can be done via using technologies that allow the device to bind between the key pair and the device itself (CA) without a possibility to expose the private key to an untrusted environment. 
		However, existing implementations only support a classic set of cryptographic operations and primitives such as AES256 or RSA. 
		It is required to develop special software for the trusted execution environment., which will support post-quantum algorithms.

		\item Assume that the set of used cryptographic algorithms and protocols is unified.
		Then the authentication of parties and verification processes are also unified. 
		This assumption is applicable to both the production line and the endpoint device itself. 
		It is important to keep in-line both software and certificates on both ends. 

		\item We recommend keeping in mind the following recommendations regarding key hierarchy.

		\end{itemize}
		
	\item Certificates revocation and compromise detection.
	
		\begin{itemize}
		
		\item If the enrollment in the device certificates (or CA itself) was compromised or expired, the functionality of the device should be limited. 
		The related system should be isolated from the device itself. 
		It is hard to achieve if the device is isolated from the public network. 
		For this type of device, it is important to enforce policies regarding the lifetime of certificates.
	
		\item Revocation lists should be maintained and updated on a regular basis. 
		For offline devices, it can be delivered with firmware updates.
	
		\item We recommend developing the PKI model in such a way that allows one to precisely revoke certificates for a specific set of devices. 
		For example, if the specific model of the device is compromised, the revocation of the certificate would not affect other products.
		
		\end{itemize}
		
\end{itemize}

We place a more technical and detailed descriptions of these recommendations in Appendix B. 

\section{Appropriate post-quantum cryptographic scheme}\label{sec:schemes}

Here we discuss the applicability of post-quantum algorithms, which depends on their parameters. 
In particular, we present the results of collecting benchmarks for various post-quantum signature schemes, which can be used for deploying quantum-secured PKI. 
We use (i) security and (ii) performance (time and key sizes) of the algorithms 
All of the presented algorithms are currently in the third round of the NIST standardization process. 

For this analysis, we take algorithms with classical security on the level of about 190 bits; see Table~\ref{tab:security}.
We note that all basic mathematical approaches used in post-quantum cryptography: multivariate cryptography, zero-knowledge proof systems, cryptographic hash functions, and lattices -- are presented.

For our tests of the algorithms with respect to time and memory consumptions, we use Intel(R) Core(TM) i5-6267U CPU @ 2.90GHz, see Table~\ref{tab:timememory}.
We note that the parameters can alternate as the security level changes. 
Falcon and qTESLA demonstrate pretty good tradeoffs both in memory and time consumption. 
However, for some special cases where one is interested in the smallest public keys size or signatures size, there are more preferable variants. 
We also note that the basic mathematical approach and status of a security proof should also be considered.

\section{Conclusion}\label{sec:conc}

The impact of quantum computing is an important aspect that is analyzed account in the development of PKI systems to protect production environments.
We have analyzed the security issues of the model of injecting the trusted certificate and provide security recommendations regarding the PKI from attacks with quantum computers.
Although our main focus is on the attacks with quantum computing, we also discuss some security issues that are not related to the used cryptographic algorithms but are important for the overall security of the PKI.
Examples of such recommendation include: 
\begin{itemize}
	\item universal security requirements for the used software and SDKs; 
	\item choosing the format of certificates that support crypto-agility and hybrid schemes;
	\item limited use certificates with a very short lifetime, which are signed with the private key of the CA;
	\item the monitoring of the modern cryptography solutions concerning non-quantum attacks and to develop maintenance procedures used to migrate possible threats;
	\item enforcing the mechanisms that allow one to revoke certificates for a specific set of devices.
\end{itemize}

The central recommendation is to realize the ability to use the hybrid cryptographic schemes~\cite{Fedorov2017} using currently standartized solutions and post-quantum solutions. 
Importantly, the candidate for the post-quantum part should be chosen according to the requirements on the size of the communications and/or time.
We have also presented various benchmark post-quantum cryptographic primitives and discussed their applicability in the security systems for production environments.

\subsection*{List of abbreviations}

\smallskip

PKI -- public key infrastructure.

NP -- nondeterministic polynomial time.

RSA cryptosystem -- Rivest-Shamir-Adleman (RSA) cryptosystem.

TLS -- Transport Layer Security.

HTTPS -- Hypertext Transfer Protocol Secure.

ECDSA -- Elliptic Curve Digital Signature Algorithm.

NIST -- National Institute of Standards and Technology.

CA -- Certificate Authority.

SHA -- Secure Hash Algorithms.

AES -- Advanced Encryption Standard.

CSR -- Certificate Signing Request.

TPM -- Trusted Platform Module.

TEE -- Trusted Execution Environment. 

SDK --  Software Development Kit. 

\section*{Declarations}

\subsection*{Availability of supporting data}

The data that support the findings of this study are available from the corresponding author (A.K.F.) on reasonable request.

\subsection*{Competing interests}

Owing to the employments and consulting activities of S.E.Y., M.K., N.P., D.N., M.A., A.G., E.O.K., and A.K.F., they have financial interests in the commercial applications of quantum-secured cryptography.

\subsection*{Funding}

This work is supported by Bosch. 

\subsection*{Authors' contributions}

All the authors contributed to the analysis of the data. 
S.E.Y., M.K., N.P., D.N., and E.K. have prepared the set of recommendation. 
D.N., M.A., and E.O.K. performed benchmarking post-quatnum algorithms. 
A.K.F. and E.O.K. wrote the manuscript with significant contribution of all the authors.
A.K.F., A.G., and A.B. supervised the project.

\subsection*{Acknowledgements}
We thank Bosch for providing the PKI scheme.

\begin{table}[]
\begin{center}
\begin{tabular}{|l|l|l|l|l|}
\hline
Algorithm                 & \begin{tabular}[c]{@{}l@{}}Basic \\ approach\end{tabular}                                & \begin{tabular}[c]{@{}l@{}}Variant of \\ the algorithm\end{tabular} & \begin{tabular}[c]{@{}l@{}}Classical \\ security, bit\end{tabular} & \begin{tabular}[c]{@{}l@{}}Quantum\\ security, bit\end{tabular} \\ \hline
Falcon                    & Lattice                                                                                  & n=768                                                               & 195                                                                & 172                                                             \\ \hline
\begin{tabular}{c}CRYSTALS-\\DILITHIUM\end{tabular}        & Lattice                                                                                  & Very high                                                           & 176/174                                                            & 160/158                                                         \\ \hline
Rainbow                   & \multirow{2}{*}{Multivariate}                                                            & Classic                                                             & 207                                                                & 169                                                             \\ \cline{3-5}
                          &                                                                                          & Compressed                                                   & 207                                                                & 169                                                             \\ \hline                                                     
\multirow{3}{*}{GeMSS}    & \multirow{3}{*}{\begin{tabular}[c]{@{}l@{}}Multivariate \\ cryptography\end{tabular}}    & GeMSS192                                                            & 192                                                                & 112.2                                                           \\ \cline{3-5} 
                          &                                                                                          & BlueGeMSS192                                                        & 192                                                                & 112.2                                                           \\ \cline{3-5} 
                          &                                                                                          & RedGeMSS192                                                         & 192                                                                & 112.2                                                           \\ \hline
\multirow{3}{*}{Picnic}   & \multirow{3}{*}{\begin{tabular}[c]{@{}l@{}}Zero-knowledge \\ proof systems\end{tabular}} & picnic-L3-FS                                                        & 192                                                                & 96                                                              \\ \cline{3-5} 
                          &                                                                                          & picnic-L3-UR                                                        & 192                                                                & 96                                                              \\ \cline{3-5} 
                          &                                                                                          & picnic2-L3-FS                                                       & 192                                                                & 96                                                              \\ \hline
\multirow{3}{*}{SPHINCS$^+$} & \multirow{3}{*}{Hash functions}                                                       & sphincs-haraka-192f                                                 & 194                                                                & 97                                                              \\ \cline{3-5} 
                          &                                                                                          & sphincs-sha256-192s                                                 & 196                                                                & 98                                                              \\ \cline{3-5} 
                          &                                                                                          & sphincs-shake256-192f                                               & 194                                                                & 97                                                              
\\ \hline
\end{tabular}
\caption{Security of post-quantum digital signature schemes. Two values of security bits for CRYSTALS-DILITHIUM are given with respect to short integer solution (SIS) and learning with errors (LWE) problems, correspondingly.}
\label{tab:security}
\end{center}
\end{table}

\begin{table}[]
\begin{center}
\begin{tabular}{|l|l|l|l|l|l|l|l|}
\hline
Algorithm                 & \begin{tabular}[c]{@{}l@{}}Variant of \\ the Algorithm\end{tabular} & \begin{tabular}[c]{@{}l@{}}Key \\ generation, \\ $\mu$s\end{tabular} & \begin{tabular}[c]{@{}l@{}}Signing,  \\ $\mu$s\end{tabular} & \begin{tabular}[c]{@{}l@{}}Signature \\ verification, \\ $\mu$s\end{tabular} & \begin{tabular}[c]{@{}l@{}}Public \\ key size, \\ byte\end{tabular} & \begin{tabular}[c]{@{}l@{}}Signature \\ size, \\ byte\end{tabular} & \begin{tabular}[c]{@{}l@{}}Secret \\ key size, \\ byte\end{tabular} \\ \hline
Falcon                    & n=768                                                               & 13882                                                                & 562                                                         & 87                                                                           & 1441                                                                & 1036.02                                                            & 6145                                                                \\ \hline
\begin{tabular}{c}CRYSTALS-\\DILITHIUM\end{tabular}        & Very high                                                           & 88                                                                   & 203                                                         & 89                                                                           & 1760                                                                & 3366                                                               & 3856                                                                \\ \hline
\multirow{2}{*}{Rainbow}  & Classic                                                             & 34980                                                                & 277                                                         & 317                                                                          & 710640                                                              & 156                                                                & 511448                                                              \\ \cline{2-8}
                          & Compressed                                                   & 41371                                                                & 24707                                                       & 7094                                                                         & 206744                                                              & 156                                                                & 64                                                                  \\ \hline
\multirow{3}{*}{GeMSS}    & GeMSS192                                                            & 79817                                                                & 900851                                                      & 478                                                                          & 1304192                                                             & 52                                                                 & 40280                                                               \\ \cline{2-8} 
                          & BlueGeMSS192                                                        & 81263                                                                & 132560                                                      & 557                                                                          & 1331744                                                             & 53                                                                 & 41720                                                               \\ \cline{2-8} 
                          & RedGeMSS192                                                         & 83529                                                                & 3672                                                        & 447                                                                          & 1359584                                                             & 55                                                                 & 40760                                                               \\ \hline
\multirow{3}{*}{Picnic}   & picnic-L3-FS                                                        & 18                                                                   & 10064                                                       & 8608                                                                         & 49                                                                  & 74191.2                                                            & 73                                                                  \\ \cline{2-8} 
                          & picnic-L3-UR                                                        & 24                                                                   & 13224                                                       & 11088                                                                        & 49                                                                  & 121849                                                             & 73                                                                  \\ \cline{2-8} 
                          & picnic2-L3-FS                                                       & 20                                                                   & 443936                                                      & 157228                                                                       & 49                                                                  & 27062.15                                                           & 73                                                                  \\ \hline
\multirow{3}{*}{SPHINCS$^+$} & \begin{tabular}[c]{@{}l@{}}sphincs-\\ haraka-192f\end{tabular}      & 14844                                                                & 439211                                                      & 21963                                                                        & 48                                                                  & 35664                                                              & 96                                                                  \\ \cline{2-8} 
                          & \begin{tabular}[c]{@{}l@{}}sphincs-\\ sha256-192s\end{tabular}      & 195818                                                               & 4390120                                                     & 3486                                                                         & 48                                                                  & 17064                                                              & 96                                                                  \\ \cline{2-8} 
                          & \begin{tabular}[c]{@{}l@{}}sphincs-\\ shake256-192f\end{tabular}    & 8767                                                                 & 240173                                                      & 12405                                                                        & 48                                                                  & 35664                                                              & 96                                                                  \\ \hline
\end{tabular}
\caption{Time consumption and memory consumption of post-quantum digital signature schemes.}
\label{tab:timememory}
\end{center}
\end{table}

\section*{Appendix A. Additional assumptions in the PKI analysis}

We also would like to note that the provided scheme is based on the following additional assumptions.
\begin{itemize}
	\item The actual process of the certificate injection for both external parties (Manufacturer and Operator) is equal as both parties rely on the PKI provided by the Maintainer.
	\item The communication of devices with the PKI frontend may be limited during the production phase due to the limited time or security concerns regarding the perimeter's isolation. 
	Due to the described limitation, it may be required to set up the PKI frontend, which mirrors the functionality of the actual PKI frontend. This may be achieved with the usage of the 2nd level certificate issued by the Maintainer.
	\item
	To implement the protection of data in transit for both parties participating in the certificate injection, it is required to embed the SHA hash of the production line PKI frontend and the main PKI frontend of the Maintainer. 
	It will allow performing the certificate pinning during the TLS communication. It is also strongly recommended to use the TLS protocol version at least 1.2.
	\item
	As a part of the certificate signing request (CSR) creation, it is obligatory for the device to perform self-attestation. 
	It may be implemented with the trusted platform module (TPM) and the trusted execution environment (TEE) on the device.
	\item 
	The TPM/TEE of the device contains the private key of the Manufacturer; the public part of the key is distributed to the PKI frontend.
\end{itemize}

\section*{Appendix. Security recommendations}\label{sec:recommendations}

Here we provide a detailed list of recommendations regarding the overall structure of the PKI. 
We analyze cryptography in place and in communications, as well as cryptographic attacks. 

\subsection*{Cryptography in place}

As CAs certificates and cryptography considered to be unified, 
we additionally assume that all parties sharing the same set of software development kits (SDKs) and software/hardware to perform required cryptography operations. 
To achieve this, the first step is to enforce universal security requirements for the software. 
Additionally, such software should pass security evaluation and should be developed according to the Security Code Practice.

As SDKs in this model are assumed to be unified, it is possible to improve the security of cryptography operations. 
For example, it is possible to embed the information regarding the current state of the service zone and used software in the certificate itself to ensure that the state of the CA is trusted. 
Moreover, the time required for the migration of the architecture to the post-quantum era, in this case, is significantly reduced since one can use the unified mechanism of the software update and deployment.

We recommend using the X.509 format for the certificate. 
This due to the fact that it supports an extensible scheme of embedded data. 
It is possible to store multiple public keys from different algorithms in the same certificate. 
For example, it is possible to embed in the signed certificate both keys RSA key and post-quantum Falcon key. 
Such a hybrid approach allows both supporting existing standards in cryptography and ensuring post-quantum security. 
However, the rollback protection mechanism must be implemented and enforced to mitigate downgrade attacks against the proposed hybrid scheme.

\subsection*{Communications}

The communication of involved parties considered to be going over the TLS connection. 
As the algorithm allows one to communicate certificate pinning for both parties, it is possible to implement a mutual authentication for involved parties. 
However, the TLS protocol by itself is not able to provide neither the integrity nor uniqueness of the data going through the tunnel. 
Moreover, as from the perspective of the public network the data itself is not encrypted from the perspective of cryptography as the TLS protocol is supposed to be used only as a way to perform the mutual authentication of parties.

As parties during the communication can operate in different time zones and conditions, it is possible for one of the parties to be unavailable during the required time period. 
A presumable solution for such a challenge is to use limited use certificates with a very short lifetime, which are signed with the private key of the CA. 
In that case, parties are able to exchange required cryptographic data for a specific set of tasks, which should be done in the near future. 
The actual confirmation from the involved parties can be received lately. 
If the certification process and signing keys at some point become compromised, then it is possible to revoke specific sets of certificates without the affection of the overall certificate chain.
This is important for the continuation of the production processes. 

{\bf Cryptography attacks}.
Cryptography plays a central role in the mentioned processes. 
However, at some point, the used cryptography tools may become vulnerable due to finding new attacks against specific modes of cryptography algorithms or due to the significant breakout in cryptoanalysis. 
For example, multiple algorithms were broken and found vulnerable due to the increased computation speed (e.g. Digital Encryption Standard). 
However, modern algorithms use the key length, which is resistance against sizeable achievements with respect to solving computational problems.

At the same time, asymmetric algorithms are much more tricky in their implementation. 
They are usually based on the assumption that a specific set of mathematical operations is hardly possible to be inverted. 
Wrong or improper optimizations of such algorithms may lead to the massive compromise of private keys. 
Examples of such drawbacks include Coppersmith's attack against the RSA algorithm~\cite{Coppersmith1996}, which is caused by the weak exponent that is used to speed up the computation of keys. 
Then one of the main recommendations is to perform the monitoring of the modern cryptography solutions and to develop maintenance procedures used to mitigate possible threats.
For the maintenance and development team, it is crucial to follow established procedures during the initial phase of the project routines, required to mitigate possible security breaches caused by modern attacks against classic cryptography.

As it is mentioned, the security of asymmetric keys based on some assumptions on the computational complexity of some mathematical problems (see above). 
Attacks with quantum computers are able to completely compromise the PKI model that is based on the usage of a set of algorithms, which are not resistant to quantum attacks. 
The extensible scheme, which allows one to replaces signing algorithms on-a-fly requires significant changes in the manufacturing cycle (e.g. firmware verification, secure boot, certificates enrollment).

As an additional improvement, it is recommended to develop the PKI model with the possibility to extend a set of used algorithms with the support of post-quantum algorithms and to perform a regular evaluation of the implemented scheme. 
It should be ensured that the scheme works in a crypto-agile manner. 
This means that tools support the replacement of algorithms on-the-fly without a significant downgrade of the scheme performance and reflection on the production line. 
In addition, it is possible to develop the PKI model using a hybrid approach (see above), which allows switching between classic and post-quantum algorithms at the authority side. 
While certificates themselves can be signed both by classical and post-quantum secure algorithms. 

\subsection*{Enrollment and provision of certificates}

The enrollment process is the initial point of the PKI model, so it deserves additional attention before the process of certificate generation can be started. 
As a consequence, the PKI model should include the trusted channel between parties, which allows parties to ensure their states and initializing the enrollment process. 
In a previous section, it is mentioned that all communication between parties should be conducted over the mutually authenticated channel.
However, we do not recommend using the same channel both for the exchange of certificates (cryptographic materials) and control signals. 

Additionally, we recommend using hardware-backed authentication methods for the critical parts of the enrollment process (e.g. confirmation of the signing of the second level certificate). 
This can be done with the help of USB tokens or similar solutions.

By taking future steps, it becomes possible also to improve the trustworthiness of CAs. 
This can be done via using technologies that allow the device to bind between the key pair and the device itself (CA) without a possibility to expose the private key to an untrusted environment. 
Moreover, depending on the used implementation it is possible to perform secure key wrapping for symmetric and asymmetric keys in a way that allows transferring keys over insecure channels, which are encrypted with the public key from the Trust Zone. 
The usage of the symmetric keys allows adding the encryption layer to the communication channel between parties. 
However, existing implementations only support a classic set of cryptographic operations and primitives such as AES256 or RSA. 
It is required to develop special software for the TEE, which will support post-quantum algorithms.

{\bf Authentication of parties and certificates verification}.
Assume that the set of used cryptographic algorithms and protocols is unified.
Then the authentication of parties and verification processes are also unified. 
This assumption is applicable to both the production line and the endpoint device itself. 
As a consequence, it is important to keep in-line both software and certificates on both ends. 
As a matter of this paper to provide recommendations regarding key hierarchy, details regarding the process of the Over-the-Air (OTA) updates and CAs themselves are considered to be outside of the context of this document. 

We recommend keeping in mind the following recommendations regarding key hierarchy.
\begin{itemize}
	\item 
	The verification process should involve an Access Control List (ACL) to limit the access granted to involved parties. 
	For example, revocation lists must be signed with the root CA certificate itself.
	\item
	The certificate itself can be bound with the device itself. 
	For example, during the communication with the backend, the device can provide unique identifiers of connected peripheral components along with the unique challenge provided by the backend. 
	This information may be used as a part of the attestation certificate provided by the device.
	\item
	The certificate itself can be collapsed. 
	For example, the device may ask the root CA to provide a new certificate using the third-level certificate issued by the manufacturer as evidence. 
	For specific cases, this functionality may reduce the complexity of the overall system. 
	Moreover, it allows implementing a flexible scheme for the usage of short life certificates.
	\item
	Runtime environment and used cryptographic software must be up-to-date (e.g., TLS protocol version and its implementations).
\end{itemize}

\subsection*{Certificates revocation and compromise detection}

Remind that the certificate revocation is a process, which can be performed both as a part of regular activities (certificate expiration) or due to the compromise.
\begin{itemize}
	\item
	If the enrollment in the device certificates (or CA itself) was compromised or expired, the functionality of the device should be limited. 
	The related system should be isolated from the device itself. 
	It is hard to achieve if the device is isolated from the public network. 
	For this type of device, it is important to enforce policies regarding the lifetime of certificates.
	\item
	Revocation lists should be maintained and updated on a regular basis. 
	For offline devices, it can be delivered with firmware updates.
\end{itemize}

Thus, we recommend developing the PKI model in such a way that allows one to precisely revoke certificates for a specific set of devices. 
For example, if the specific model of the device is compromised, the revocation of the certificate would not affect other products.

\subsection*{Symmetric key server}

As an alternative for the implementation of a system aiming to protect intellectual property and authentication of parties, 
it is possible to integrate into the production environment a key server.
Key servers perform the authentication of parties using a stored list of hashes of passwords. 

In the current state of business requirements regarding the injection of certificates, 
it is hardly possible to use the plain key server implementation for each involved party. 
However, it may be possible to improve the introduced scheme of the PKI with some elements of the symmetric key server.
We note that symmetric cryptographic algorithms are considered resistant against the attacks with quantum computers 
(under the condition that the key distribution process is also based on quantum-secured schemes).

For example, production line servers, which inject the operator certificate, may be authorized with credentials to the main PKI frontend to provide the current state of the certificate enrollment. 
As it may be hard to enforce the usage of the unified backend/software for the production environment, the implementation of the authentication of manufacturers based on the credentials may allow mitigating some problems.
For example, the manufacturer may use the key server to provide the information regarding injected certificates to the PKI frontend. 
Consequently, the PKI frontend may use this information during the enrollment of the operator certificate. 
In that way, the PKI environment of the manufacturer may be completely isolated from the maintainer.

The scalability of the key server is usually quite limited as it requires storing a significant amount of data and processing a large amount of requests 24/7. 
However, it may be possible to introduce it as a part of the PKI model to mitigate some bottlenecks. 

\end{document}